\documentclass[mathleft
]{an}
\usepackage{graphicx}
\usepackage{times}
\begin{document}

\Pagespan{433}{437}
\Yearpublication{2012}%
\Yearsubmission{2012}%
\Month{06}%
\Volume{333}%
\Issue{5}%
\DOI{10.1002/asna.201211682}

\title{Cosmic Flows surveys and CLUES simulations}

\author{H.M. Courtois\inst{1,2}\fnmsep\thanks{Corresponding author:
  \email{h.courtois@ipnl.in2p3.fr}\newline}
\and  R.B. Tully\inst{2}
}
\titlerunning{Cosmic Flows}
\authorrunning{H.M. Courtois, B. Tully}
\institute{
University of Lyon; UCB Lyon 1/CNRS/IN2P3; IPNL, 4 rue Enrico Fermi, 69622 Villeurbanne, France
\and 
Institute for Astronomy, University of Hawaii, 2680 Woodlawn Drive, Honolulu, HI 96822, USA}

\received{2012 Jan 17}
\accepted{2012 Mar 12}
\publonline{2012 Jun 1}


\abstract{%
 ÒCosmic FlowsÓ is a program to determine galaxy distances for 30,000 galaxies with systematic errors below 2\%, almost ten times the number currently known and a five-fold improvement in systematics. The resultant velocity field will provide input for constrained local universe simulations: ÒCLUESÓ (www.clues-project.org). The observed and the simulated universe are then comparatively studied. This synergy of observations and theory distinguishes the program, and should lead to fundamental discoveries regarding the sources of deviations from the expansion of the universe. Specifically, the program should give a definitive answer to one of the most outstanding unsolved problem in cosmology: the cause of the motion of 630 km/s of our Galaxy manifested in the microwave background dipole. This paper presents current results with particular emphasis on the "great attractor" reconstruction.}

\maketitle

\section{Introduction}
The universe has assembled itself into a complex web over the course of 14 billion years.  Structure has formed due to gravitational collapse driven by dark matter.  It appears that today repulsive dark energy has gained ascendancy over dark matter, leading to the termination of collapse on large scales.  Yet we have little understanding of the nature of dark matter or dark energy, the two dominant constituents of the universe.  The goal of this program is to contribute to an improved and accurate understanding of dark matter and dark energy.

On scales of groups and clusters, and of their constituent galaxies, stars, and planets, collapse has occurred and a memory of initial conditions has largely been erased by dissipative processes.  However, on scales larger than groups the primordial contest between cosmic expansion and gravitational collapse is still being played out, modulated by a reappearance of expansion in the vacuum.  Dark matter manifests itself through deviations from the universal expansion.  We know with precision that deviations can be important: the dipole temperature variation in the Cosmic Microwave Background informs us that our Galaxy has a deviant motion of 630 km/s in a well defined direction (Fixsen et al. 1996, ApJ, 473, 576).  The preferred assumption is that we are being attracted by a lot of dark matter, but we do not know by how much, and how far away it is.  The cause of our deviant motion is an outstanding problem in cosmology. There can be no pretence that we understand the composition of the universe if we cannot plausibly explain our peculiar motion.

The distribution of dark matter can be inferred if we have a detailed map of the departures from cosmic expansion.  The radial component of the `peculiarÕ velocity of an individual galaxy, $V_{pec}$, can be determined if we know a good distance, $d$, for the galaxy since the (easily established) observed velocity of the galaxy, Vobs, is the sum of the peculiar velocity and a cosmic expansion component, $H_{0} * d$ where $H_{0}$ is the Hubble Constant: $V_{obs} = H_{0} * d + V_{pec}$.  The Hubble Constant can be established by averaging over a sufficiently large and distributed sample with distances for which $<Vpec> =0$.

There was community enthusiasm for peculiar velocity studies during the 1990Õs reaching a crest with the Cosmic Flows workshop in Victoria, Canada (Courteau et al. eds. 2000, Cosmic Flows Workshop).  There has been a hiatus since because it came to be appreciated that a lot more and better data were needed.  The case will be made here that vastly superior data are at hand and, moreover, that more sophisticated analytic tools are available.  We intend to be on the vanguard of a resurgence of activity in cosmic flows studies.

The main observational difficulty is with the measurement of galaxy distances.  Say a distance is measured with an rms uncertainty of 15\%.  For a galaxy with $V_{obs}$ = 4,000 km/s the uncertainty of the cosmic expansion component (hence the uncertainty in $V_{pec}$) will be 600 km/s, comparable to our bulk motion with respect to the Cosmic Microwave Background. Nearby we do well.  For a galaxy at 600 km/s the error is only 90 km/s.  At 10,000 km/s the error is 1,500 km/s.  Peculiar velocities of individual galaxies can be mapped in great detail nearby but at larger distances we only get statistical information.  The numerical orbit reconstruction and simulations that will be discussed play a critical intermediary role in the interpretation of line-of-sight peculiar velocities in terms of 3D motions.  Maximum likelihood models can be fit that tailor to low errors nearby and in high density regions and high errors at a distance and where data are sparse.

Nowadays there are several good ways to measure galaxy distances but there are limitations to each method.  We will incorporate all methods that are considered appropriate although our observational efforts will focus on the only method that can provide a dense grid of measurements over a large local volume even in low density regions.  Methodologies we will incorporate include those based on the Cepheid Period-Luminosity Relation (Freedman et al. 2001, ApJ, 553, 47), the Tip of the Red Giant Branch (Rizzi et al. 2007, ApJ, 661, 815), Surface Brightness Fluctuations (Tonry et al. 2001, ApJ, 546, 681), the Fundamental Plane (Colless et al. 2001, MNRAS, 321, 277), and Type Ia Supernovae (Tonry et al. 2003, ApJ, 594, 1).   These procedures can give distances to individual galaxies or groups with accuracies of 5-10\%.  However, either they can only be applied to very nearby galaxies, or galaxies of early morphological type, or serendipitously.  Our observational campaign is directed at application of the Tully-Fisher Relation (Tully \& Fisher 1977, A\&A, 54, 661), the correlation between the luminosity of a galaxy and its rotation rate.  This methodology gives individual distance measures accurate to 15\%, less precise than others but it is applicable to most spiral galaxies, the most common and widely distributed kind of luminous galaxy.  Our program will measure accurate distances to 30,000 galaxies: an order of magnitude increase over the present state-of-the-art.

A measurement of a distance with the Tully-Fisher Relation requires two observations: one to provide the luminosity of old stars, inferred to be correlated with the total mass, and one to determine the rotation of the galaxy, a kinematic response to the total mass.  Distance is given by the offset between the intrinsic luminosity gauged from the rotation rate and the observed luminosity.  While the methodology has a long history, several game-changing advances are about to be realized affecting both measurables.

\section{Radioastronomy surveys}
Galaxy rotation can most accurately be determined by observing the HI radio 21cm line.  The important new and anticipated capabilities are provided by the upgraded Arecibo Telescope in Puerto Rico, the 100 m Green Bank Telescope in the USA, ASKAP, the Australian Square Kilometre Array Pathfinder and, ultimately, the wide field-of-view upgrade to the Westerbork Synthesis Radio Telescope, Apertif, in the Netherlands.  We have been engaged in observations with the Arecibo and Parkes telescopes and, most importantly, with a Large Program on the Green Bank Telescope (http://ifa.hawaii.edu/cosmicflows).  As a result we have a well established data analysis pipeline and, with the inclusion of archival material, have assembled a database of 16,000 HI profiles (Courtois et al. 2009, AJ, 138, 1938); see http://edd.ifa.hawaii.edu and select the `All Digital HIÕ catalog. The next giant step will be with ASKAP with observations beginning in 2011.  It is anticipated to detect 200,000 extragalactic HI sources within z = 0.23 over three-quarters of the sky.  We will use our innovative newly established procedures to determine rotation rates for all adequately detected sources (8 sigma detections of the peaks).  The final quarter of the sky will be observed with Apertif (Westerbrok Netherlands), providing galaxy rotation information for 30,000 galaxies, with high density coverage within 10,000 km/s and some information as deep as 20,000 km/s. 

\begin{figure}
\includegraphics[width=9cm]{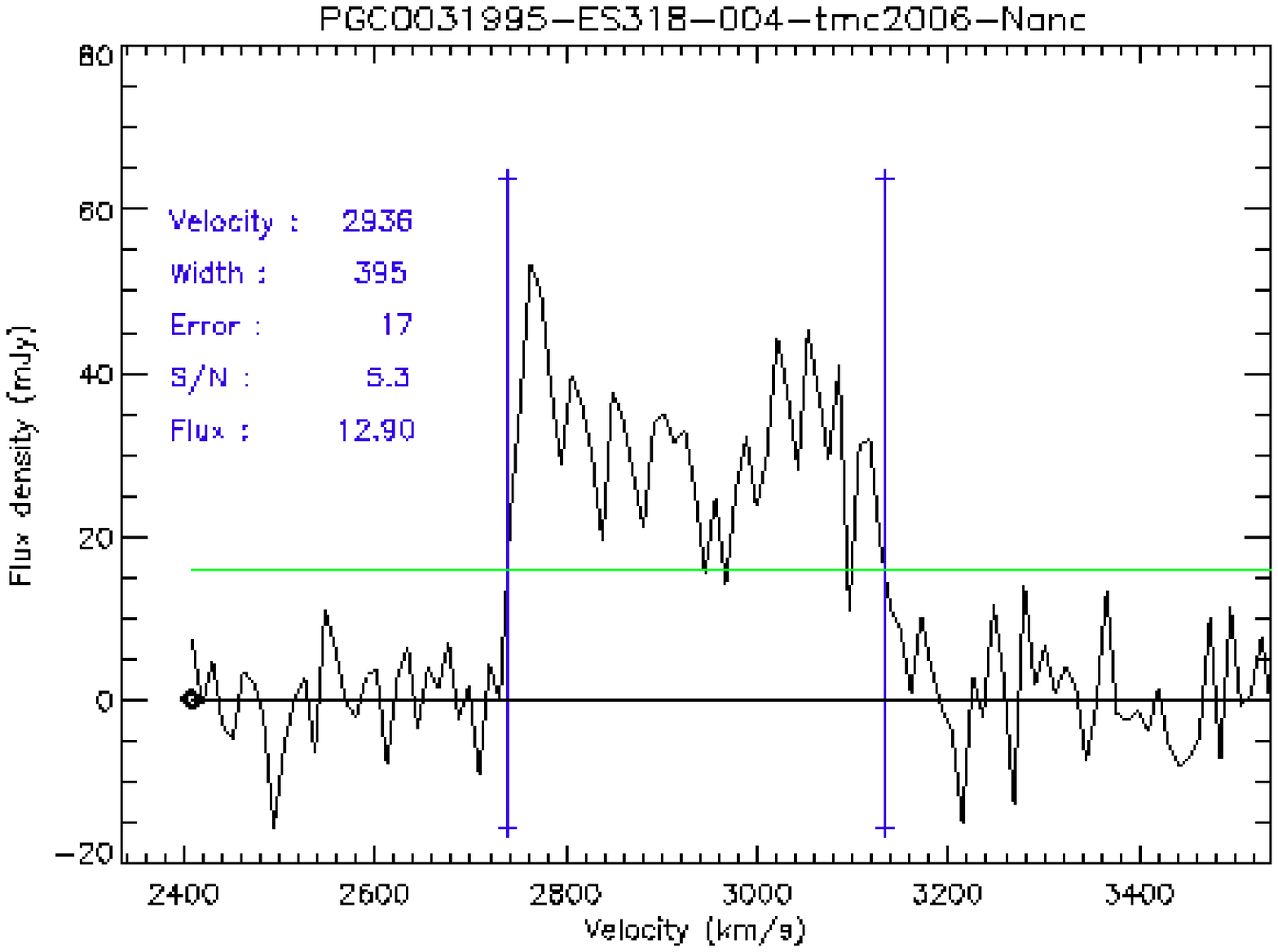}\\
\includegraphics[width=8cm]{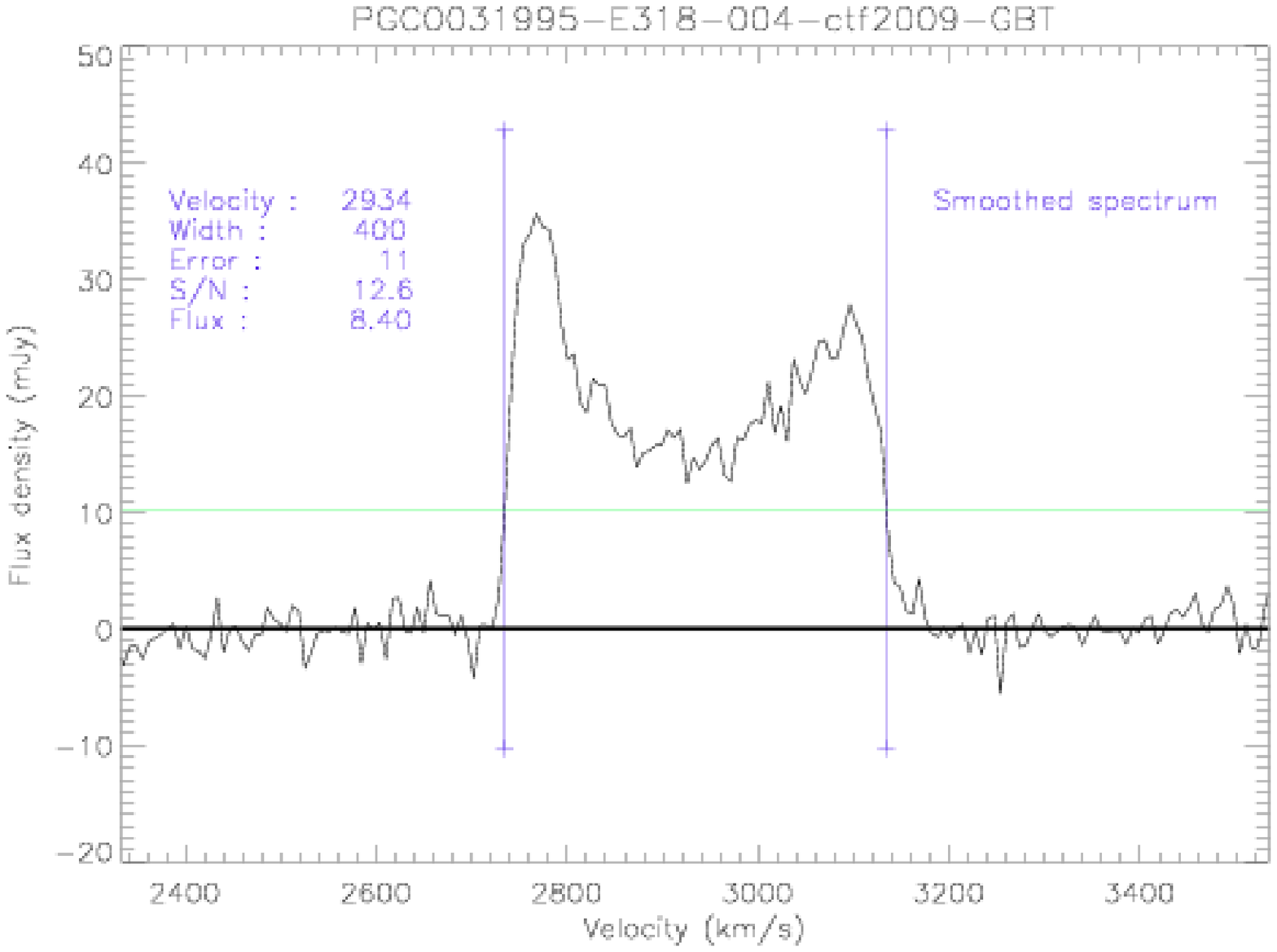}\\
\caption{The HI profile for PGC31995 as previously observed at Nancay (top) and as now obtained with the Cosmic Flows Large Program on the Green Bank Telescope (bottom).}
\label{HIprofile}
\end{figure}

\section{Surface photometry surveys}
In the past, the photometry necessary to complement the kinematic measurements have been carried out at optical and near infrared bands by a large number of players using many telescope, detector, and filter combinations in conditions with uncertain controls.  The results have been unsatisfactory. A simple thought experiment illustrates the problems experienced with photometry in the past.  Suppose there is a systematic inconsistency of 10\% between north and south or spring and fall or one literature source and another.  Such a plausible inconsistency would act on targets at 4,000 km/s to generate a false ÔflowÕ of 400 km/s.  Our expectation is that the new optical and infrared surveys will enable systematics to be held below 2\%.

The way forward has been demonstrated by two major projects.  The Two Micron All Sky Survey (2MASS) resulted in a catalog of all galaxies brighter than Ks = 13.5 (Jarrett et al. 2000, AJ, 119, 2498).  The Sloan Digital Sky Survey (SDSS) observed the northern polar cap in 4 optical bands and likewise resulted in a catalog of galaxies to faint magnitudes (http://www.sdss.org).  In both cases, the photometric integrity for point sources is outstanding.  However in the case of 2MASS the exposure times were short and the sky background at 2 microns is intense with the consequence that only the highest surface brightness components of galaxies are registered.  In the case of SDSS, the pipeline processing had poor control over sky levels with large galaxies.  It is possible to reprocess and recover adequate photometry for big galaxies on a case-by-case basis but, still, SDSS only covers a quarter of the sky.

We are involved with two alternative approaches to the photometry problem Ð either one providing an adequate solution.  The first approach derives from two ground-based optical surveys, PanSTARRS in Hawaii (http://pan-starrs.ifa.hawaii.edu) and SkyMapper in Australia (http://msowww.anu.edu.au/skymapper) which between them are covering the entire sky in g,r,i,z bands.  Both surveys are just beginning and over the course of 5 years will generate increasingly deep images of the entire sky with rigorous photometric quality control.  The second solution involves imaging at 3.6 microns from space with either of two satellites: Spitzer and WISE.  In the case of Spitzer, observations are made in pointing mode and already 2000 galaxies have been targeted and observations have been granted (200 hrs) from 2011. We are also members of the consortium S4G (Kartik Seth) who have already observed 3000 TF galaxies.  On the other hand WISE, the Wide-Field Infrared Survey Explorer, has now completed imaging of the entire sky.  Sky background which plagues ground observations in the infrared is not a problem from space.  Spitzer data are extremely deep and robust for our purposes.  WISE data are less deep but adequate.  In the case of WISE, there is not yet an extended source processing pipeline so there is work currently done in this direction.

\begin{figure}
\includegraphics[width=8cm]{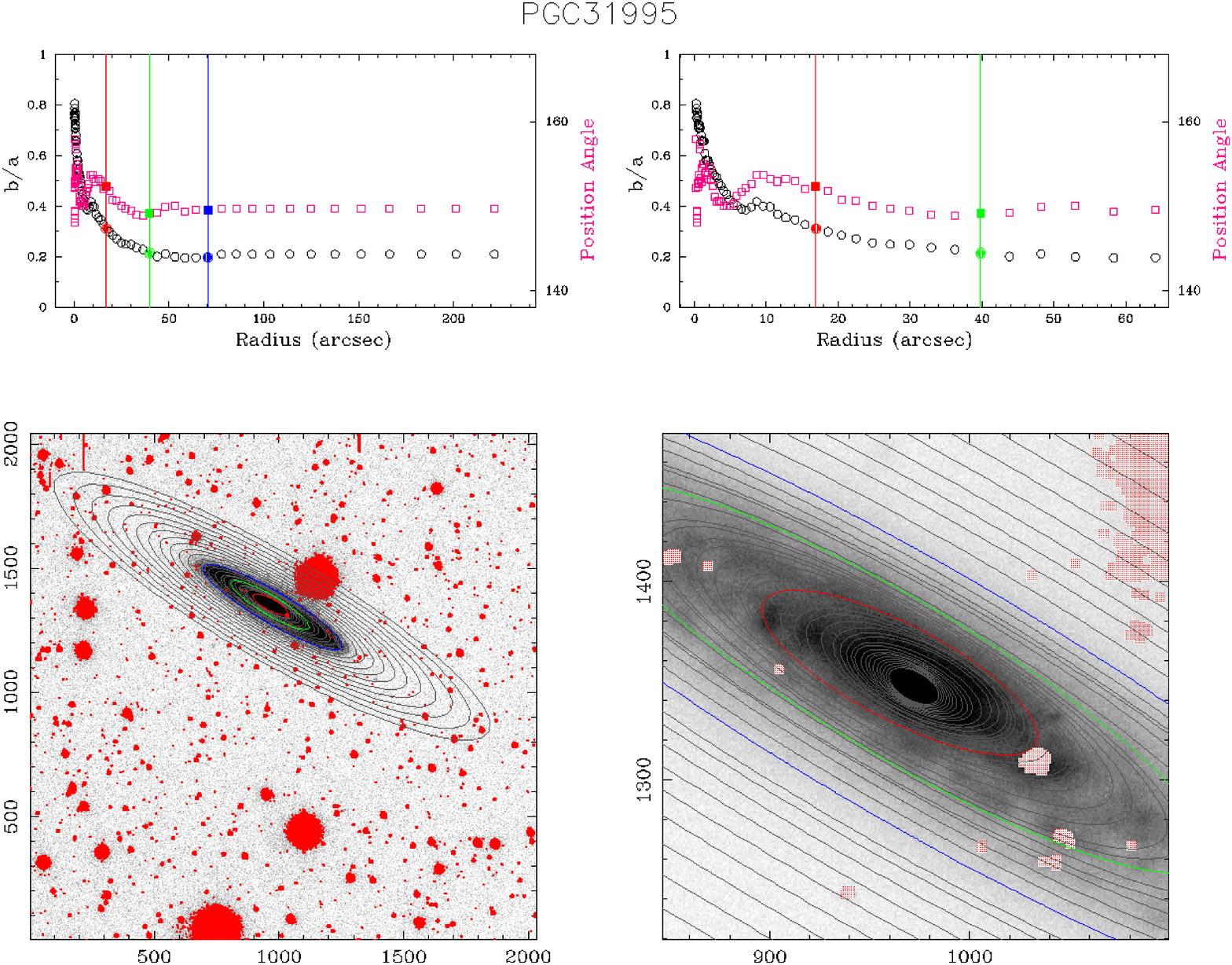}\\
\includegraphics[width=8cm]{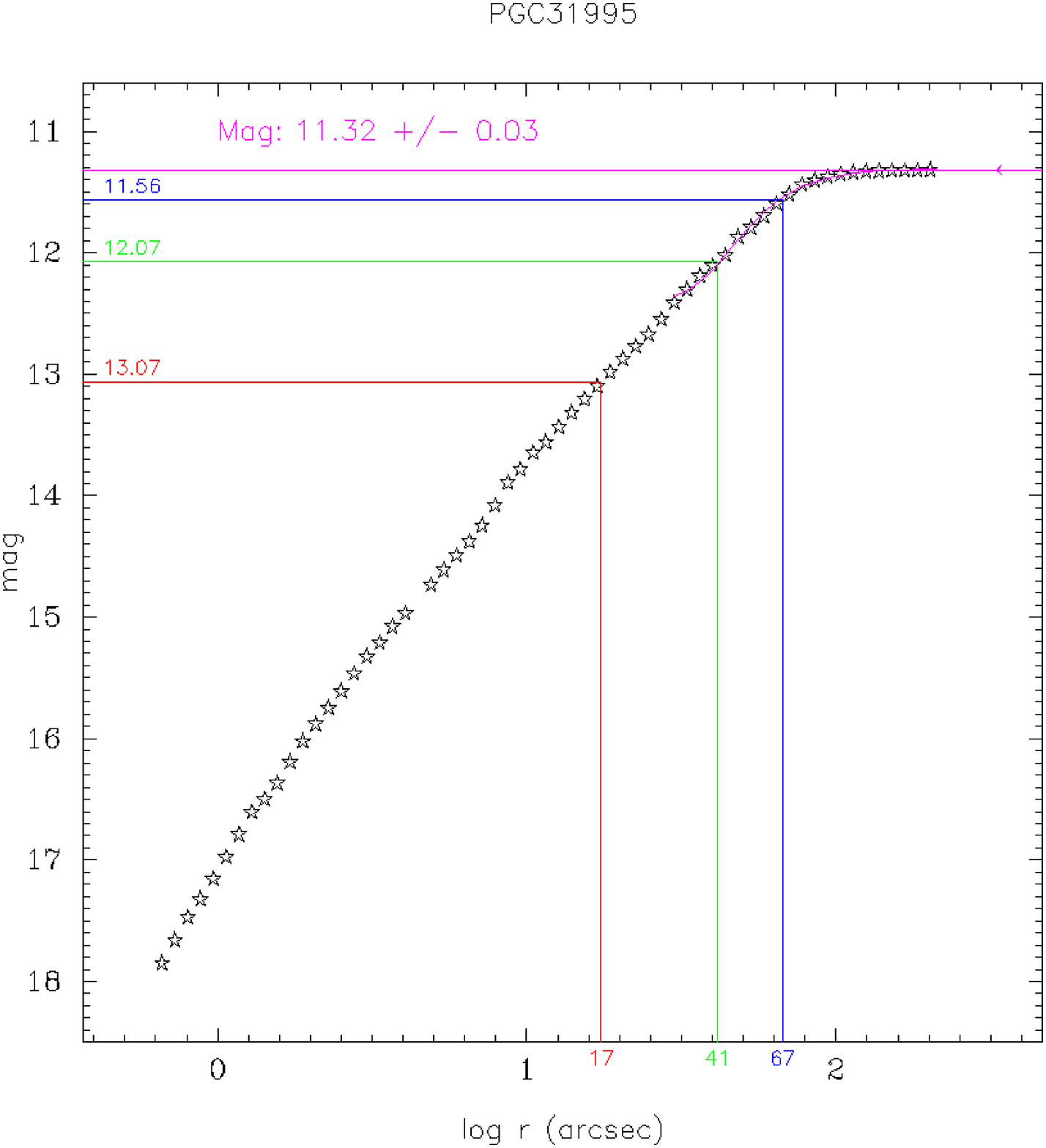}\\
\caption{Photometry results from University of Hawaii 2.2m observations of the galaxy PGC31995 (whose HI profile is illustrated in Fig~\ref{HIprofile}).  Anti-clockwise from lower left: image of the galaxy with stars masked and isophotal contours superimposed; blow-up of  center; surface brightness as a function of radius; two panels on different scales of ellipticity and position angle.}
\label{photom}
\end{figure}

\section{Databases}
The material from the large programs that have been described and more need to be managed in ways that go beyond the requirements of the individual large radio, infrared, and optical surveys described above.  We have experience with the co-management of an important extragalactic database at the University of Lyon : HyperLeda (http://leda.univ-lyon1.fr). The new database we are developing (Tully et al. 2009, AJ, 138, 323) is called EDD, the Extragalactic Distance Database (http://edd.ifa.hawaii.edu).  It serves three primary functions.  First, it provides a repository for data that we have obtained through observations or extracted from archives and reprocessed.  The most consequential current asset in EDD is the `All Digital HIÕ catalog of 16,000 consistently processed HI profiles for 14,000 galaxies (Courtois et al. 2009, AJ, 138, 1938).

The second primary function of EDD is to assemble the many distance scale building blocks so they can be evaluated and linked.  The database consists of catalogs, some extracted from the literature and some constructed by us, but in every case the line entries pertain to a specific galaxy identified by its 7 digit PGC name (Paturel et al. 1996, ISBN 2-908288-08-7).  Presently there are information on 50,000 galaxies. EDD is open without password. 

The third primary function of EDD is to facilitate the generation of three-dimensional spatial and velocity field maps of the nearby universe.  The observed building blocks must be converted into distances, alternative distances by different methodologies must be evaluated for consistency, alternative distances to the same galaxy must be reconciled, the grouping properties of galaxies is to be understood and mean distances to groups are to be found.  Then a value of the Hubble Constant consistent with the distance scaling should be determined, hopefully derived by averaging over a large enough volume to be representative of the global value.  With all this information, line-of-sight peculiar velocities can be separated from observed velocities.  

\section{Constrained cosmological simulations}
Some of our initial experiences have been with two large scale simulation projects LEGO (http://graal.ens-lyon.fr/LEGO) and `HORIZONÕ (http://project-Horizon.fr).  Progress were made by the development of the skeleton technique to demonstrate large scale connectivity (Sousbie et al. 2008, ApJ, 672, L1), the development of procedures for the generation of realistic mock catalogs (Sousbie et al. 2008, ApJ, 678, 569), and the development of a web interface to launch large cosmological simulations on a grid of computers in France (Depardon et al.  2010, AIPC, 1241, 816).

Most recently, we have joined the larger international collaboration CLUES, Constrained Local UniversE Simulations (http://www.clues-project.org) with four PIs: S. Gottloeber in Germany, Y. Hoffman in Israel, A. Klypin  in USA, and G. Yepes  in Spain (Gottloeber et al. 2010, arXiv:1005.2687).  The collaboration is generating a map of structure within 10,000 km/s where the input is the observed peculiar velocity field, the response to the distribution of matter.  Peculiar velocities and density perturbations are directly related in linear theory.  Wiener filtering (Hoffman 2009, LNP, 665, 565) of the observed velocity distribution recovers the 3-dimensional density distribution.  Observers in the collaboration provide the observational input to the models.  The task not only involves providing the velocity field information described above.  It also requires making the translation between what observers see and modelers create.  Observers see sparsely distributed galaxies but only the most luminous ones at larger distances and not those in regions of obscuration.  Modelers create dense webs of particles unencumbered by losses with distance or a zone of obscuration but the interpretation of those particles in terms of galaxies is an art.

Alternative approaches are explored, they are based on the reconstruction of galaxy orbits to todayÕs observed positions and velocities from initial conditions that globally minimize the action.  One procedure developed by colleagues in France that goes by the acronym MAK, Monge-Ampere-Kantorovich (Lavaux et al. 2010, ApJ, 709, 483), finds straight-line orbits for very large samples.  This method can be used to roughly determine the initial conditions of redshift surveys of hundreds of thousands of galaxies.  A second procedure called NAM, Numerical Action Methods (Peebles et al. 2001, ApJ, 554, 104), is restricted in use to smaller samples of order 1000 objects but is able to find physically meaningful orbits in the non-linear regimes of high overdensity with respect to the universal mean.  These several methodologies are complementary.  MAK is useful to develop a coarse picture on a large scale.  NAM is useful on small scales where detailed information is available.  It can be embedded in a MAK reconstruction.  The combination can constrain the initial density perturbation conditions locally.  The Wiener filter approach of CLUES provides an alternative route to a description of the initial conditions.  A very attractive feature of the Wiener filter approach is that a density map can be constructed by extrapolation to of order twice the dimensions (ten times the volume) of the domain of distance measurements because of coherence in the properties of flows on large scales. The initial conditions that can be recovered can serve as the starting point of a constrained N-body and hydrodynamic simulation, the results of which are to be compared with the observed world.

\section{Current results}

For our first generation of results from the "cosmic flows" program, we used our currently available catalog of distance measurements containing 1,797 galaxies within 3000 km/s: Cosmicflows-1.  The Wiener Filter method was used to recover the full 3D peculiar velocity field from the observed map of radial velocities and to recover the underlying linear density field (Courtois, Hoffman  et al. 2012 ApJ 744, 43). The velocity field within a data zone of 3000 km/s was decomposed into a local component that is generated within the data zone and a tidal one that is generated by the mass distribution outside that zone.  The tidal component is characterized by a coherent flow toward the Norma-Hydra-Centaurus (Great Attractor) region while the local component is dominated by a flow toward the Virgo Cluster and away from the Local Void.  A detailed analysis shows that the local flow is predominantly governed by the Local Void and the Virgo Cluster plays a lesser role.   

The motion identified as `Local' is in excellent agreement in both amplitude and direction with the local motion identified by Tully et al.  2008 (a motion of the Local Sheet of $323 \pm 25$ km/s toward $l_{gal} = 220 \pm 7^\circ$, $b_{gal} = +32 \pm 6^\circ$ or $sgl = 80^\circ$, $sgb = -52^\circ$).  The dominance of the Local Void marginalizes the dynamical role of the Virgo cluster on the local scene. 
The motion of the Local Group with respect to the CMB is recovered by the present reconstruction to within 5\% in amplitude and 5 degrees in direction. 

Decomposing the velocity field into local and tidal components, with respect to the data zone sphere of R=3000 $Mpc.h{^-1}$, we find that the local structure contributes a component of  304 km/s  leaving a residual tidal component of  382 km/s, both deviating from the CMB dipole by $30^\circ$ but  in almost orthogonal directions. 

\begin{figure}
\includegraphics[width=8cm]{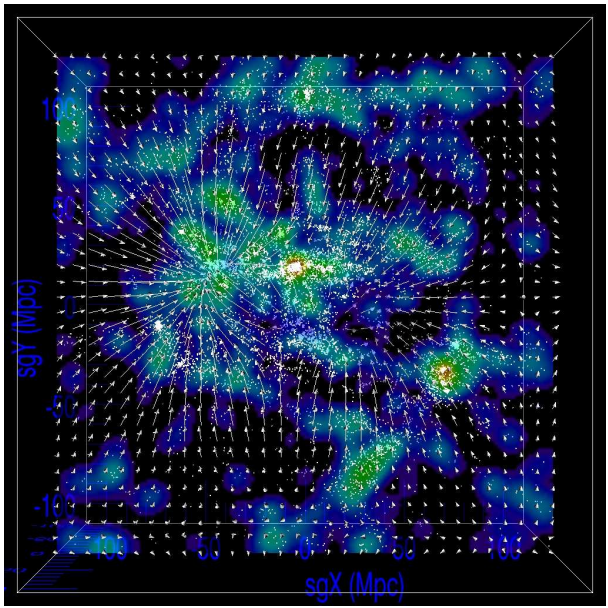}\\
\caption{3D reconstruction using the Wiener Filter method. The great attractor infall is seen on the center left of the image.}
\label{wienerfilter}
\end{figure}

Additionally a  method is being developped to reconstruct the cosmological displacement
field using the peculiar velocity of dark matter haloes at z = 0 as a proxy. We apply
the Zel'dovich approximation in reverse to link the velocity and displacement  fields and to trace the haloes back to their initial position. We quantify the error of this procedure depending on different properties of the haloes and their environment (Doumler et al. 2012 in prep.). We then investigate how this method can be applied to realistic radial peculiar velocity data.

Cosmologogical simulations using the catalog cosmicflows-1 as initial conditions will be the second generation of the CLUES constrained simulations.

\section{Conclusions}

Peculiar velocities are signatures of the distribution of dark and luminous matter.  Collapse timing is a signature of dark energy.  Velocity field studies are complementary to weak lensing, the only alternative method for directly measuring the amplitude of dark matter fluctuations on scales that have not yet collapsed.  Weak lensing provides statistical information on relatively distant structures (Kaiser \& Squires 1993, ApJ, 404, 441) while velocity field maps provide information that is fine-grained locally and degrades with distance.

On the observational side the new capabilities for spectroscopy in the HI line with the new radio telescopes and photometry with all-sky surveys from the ground at optical wavelengths and from satellites in the infrared are transformative.  We can now envisage a sample for distances of 30,000 galaxies with integrity in the photometric calibration at the level of 2\%.   This is a very high gain over the existing situation (of order 4,000 galaxies with some sort of distance measured and with systematics at the level of 10\%).  

Through a convergence of observational opportunities and computational and display capabilities it is now possible to make serious progress with one of the outstanding problems in cosmology: the nature of deviant galaxy motions Ð presumably a response to the properties of the dark sector.  The deviant motions are seen over a wide range of scales.  Locally, random motions are seen to be remarkably small but there can be abrupt discontinuities.  On large scales, coherent flows of impressively large amplitude are seen.  There is still no agreement on the scales and amplitudes.  There are hints that these flows are larger than can be understood by the standard Lambda Cold Dark Matter paradigm. 
Researchers and the public are puzzled by the legitimate question: Where are the Dark Energy and Dark Matter that dominate the cosmos?

\end{document}